%Paper: hep-ph/9303203
%From: U6404939@ucsvc.ucs.unimelb.edu.au
%Date: 02 Mar 1993 13:14:42 +1100

\magnification=1200
 {\nopagenumbers
\baselineskip=15pt
\hsize=5.5in
\hskip10cm UM--P-93/16\hfil\par
\hskip10cm OZ-93/6 \hfil
\vskip2cm
\centerline{\bf CP Test in the W Pair Production} \par
\centerline{\bf via Photon Fusion at NLC}\par
\vskip2cm
\hskip4cm J.P. Ma and B.H.J. McKellar \par
\vskip0.5cm
\hskip4cm Recearch Center for High Energy Physics \par
\hskip4cm School of Physics\par
\hskip4cm University of Melbourne \par
\hskip4cm Parkville, Victoria 3052\par
\hskip4cm Australia \par
\vskip2cm
{\bf\underbar{Abstract}}:\par
We study the possibility to test CP invariance in the $W^+W^-$ production
via photon fusion at NLC. The predictions of the CP violation effects
are made within two Higgs doublet extensions of the minimal standard
model, where CP violation is introduced by a neutral Higgs exchange in
s channel in our case. The width effect in the Higgs propagator on the
CP violation effects is studied in detail. The CP violation effects
can be measured in some parameter region of the extensions. \par
\par\vfil\eject }
\baselineskip=20pt
\pageno=1
\noindent {\bf 1. Introduction}\par
The $W$ pair production at $e^+e^-$ colliders is an impotant process, in which
the interactions between the weak gauge bosons in the Standard Model can
be studied. But, as the c.m.s. energy of a $e^+e^-$ collider increases, the
cross section for $e^++e^-\rightarrow W^++W^-$ decreases
as expected from gauge invariance. At NLC with
$\sqrt{s}=500$GeV, the
cross section $\sigma (e^+e^-\rightarrow W^+W^-)$ is only about
7pb. The idea of using Compton(back) scattering of laser light
to obtain $\gamma\gamma$ collisions at NLC with approximately
the same energies and luminosities[1], brings new possibilities to the study
 of couplings between the gauge bosons,
because at higher energy  more W pairs are produced
via photon fusion than via $e^+e^-$ collision. This has
recently attracted attention to the study of CP conserving
anomalous couplings between the gauge
bosons in $\gamma +\gamma \rightarrow W^+ +W^-$[2,3]. In this work we study
the possibility of
testing CP invariance in W pair production via photon fusion at NLC.
\par
As is well known, CP violation was found 30 years ago, but it is still
not known which form of interaction is responsible for CP violation. The
KM matrix[4] provides one  possibility. The fact that CP violation
has been
observed only in the K meson system makes it difficult to determine the
form of the CP violating interactions.
 W pair production in $\gamma\gamma$
collisions gives a new possibility of testing CP invariance, and thus to
get more information about the interaction responsible for CP violation.
\par
In Sect. 2
we will discuss how to make CP tests in the process $\gamma +\gamma
 \rightarrow W^+ +W^-$ at NLC, where the polarization
of the initial photons is not known.
We take account of the experimental situation
and our CP odd observables are constructed in such a way
that they are directly measurable in experiment without
resonstruction
of the W rest frames and the centre of momentum frame
of the initial pair of photons.
In Sect. 3 we will consider
the particular case of two Higgs doublet models, in which CP
violation is caused by neutral Higgs exchange. In Sect. 4 we give numerical
results for the observables we proposed in Sect. 2 and a summary. \par
\vfil\eject
\noindent {\bf 2. CP Constraints and CP odd Observables}\par
We consider the following process in the c.m.s. of the initial state:
  $$\eqalign{  \gamma (p_1) +\gamma (p_2)& \rightarrow W^+(k_1) +W^-(k_2) \cr
                & {\bf p_1}+{\bf p_2}=0 \cr  }\eqno(2.1)$$
and the amplitude for the process (2.1) is:
   $$ T_{fi}=\varepsilon_{\nu_1}(p_1)\varepsilon_{\nu_2}(p_2)
      \varepsilon^*_{\mu_1}(k_1)\varepsilon^*_{\mu_2}(k_2)
    A^{\nu_1\nu_2\mu_1\mu_2}(p_1,p_2,k_1,k_2) \eqno(2.2) $$
We assume that the polarization of the initial photons is not known. Then, a
CP test in (2.1) is possible only if the polarizations of the $W^+$ or $W^-$
are observed. To obtain information about the polarizations we consider
the leptonic decay of the W bosons.
That means, we actually consider the process:
  $$ \gamma (p_1) +\gamma (p_2) \rightarrow W^+(k_1)+W^-(k_2)
   \rightarrow \ell^+(q_1)+\ell^-(q_2) +{\rm neutrinos} \eqno(2.3)$$
For the decay process $W^+(k_1)\rightarrow \ell^+(q_1)+\nu $
with a moving $W^+$
a covariant decay matrix $\rho^+_{\mu\nu}(k_1,q_1)$
 can be defined and
is normalized as:
  $$ {1\over 4\pi} (k_1^0)^{-2}
     \int {d\Omega_1 \over (1-\beta {\bf \hat k_1}\cdot
 {\bf \hat q_1})^2}\rho^+_{\mu\nu}(k_1,q_1)=(-g_{\mu\nu}+{k_{1\mu}k_{1\nu}
\over M_W^2}) \eqno(2.4) $$
where
  $$  {\bf \hat k_1}={{\bf k_1}\over \vert {\bf k_1}\vert}, \  \ \
  {\bf \hat q_1}={{\bf q _1}\over \vert {\bf q_1}\vert}, \  \ \
   \beta={\vert {\bf k_1}\vert \over k_1^0} \eqno(2.5)$$
and $\Omega_1$ is the solid angle of ${\bf q_1}$. At tree-level in the SM
the decay matrix
$\rho^+_{\mu\nu}(k_1,q_1)$ takes the form:
   $$ \rho^+_{\mu\nu}(k_1,q_1)={3\over 2}(-M_W^2g_{\mu\nu}
   +2(k_{1\mu}q_{1\nu}+k_{1\nu}q_{1\mu})-4q_{1\mu}q_{1\nu}
    +2i\varepsilon_{\mu\nu\alpha\beta}k_1^\alpha q_1^\beta) \eqno(2.6)$$
Similarly, one can also obtain the covariant decay matrix $\rho^-_{\mu\nu}
 (k_2,q_2)$ for $W^-(k_2)\rightarrow \ell^-(q_2)+\bar \nu $.
The probability for the process (2.3) may be written in terms of the
decay matrices as:
   $$ \eqalign { R({\bf p_1},{\bf k_1},{\bf q_1},{\bf q_2})=&
     {1\over 4}A^{\nu_1\nu_2\mu_1\mu_2}(p_1,p_2,k_1,k_2)
    { A^*_{\nu_1\nu_2}}^{\mu'_1\mu'_2}(p_1,p_2,k_1) \cr
    &\cdot \rho^+_{\mu_1\mu'_1}(k_1,q_1)\rho^-_{\mu_2\mu'_2}(k_2,q_2)
  \cr}  \eqno(2.7)$$
If CP invariance holds, the following relation holds for $R$
  $$ R({\bf p_1},{\bf k_1},{\bf q_1},{\bf q_2})=
   R({\bf p_1},{\bf k_1},-{\bf q_2},-{\bf q_1}). \eqno(2.8) $$
\par
The expectaction value of any observable $O$, which is a function of
${\bf p_1},{\bf k_1},{\bf q_1}$ and ${\bf q_2}$ can be obtained:
 $$ \eqalign { <O> &= {1\over N} \int f_\gamma (x_1)f_\gamma (x_2) {dx_1dx_2
  \over x_1x_2}{\beta \over 2(4\pi)^2}\int d\Omega  \cr
    & \cdot {1\over 4\pi} (k_1^0)^{-2}
     \int {d\Omega_1 \over (1-\beta {\bf \hat k_1}\cdot
 {\bf \hat q_1})^2}{1\over 4\pi} (k_1^0)^{-2}
     \int {d\Omega_2\over (1+\beta {\bf \hat k_1}\cdot
 {\bf \hat q_2})^2}  \cr
  &  \cdot O \cdot R({\bf p_1},{\bf k_1},{\bf q_1},{\bf q_2})
  \cr} \eqno(2.9) $$
Here,  $d\Omega$($d\Omega_2$) is the solid angle of ${\bf k_1}$(${\bf q_2}$).
The normalization factor
$N$ is defined so that $<1>=1$. The
distribution function $f_\gamma (x)$ gives the proportion of the photons
with the fraction $x$ of the energy carried by the electron or positron.
In the expression (2.9) any experimental cuts can be included, but for our
purpose they must be CP blind.
\par
The momenta ${\bf p_1}$, ${\bf k_1}$, ${\bf q_1}$ and ${\bf q_2}$ are not
directly measurable. Due to the missing neutrinos and
the lack of knowledge
about the c.m.s. of the initial state they can never be known completely
in the experiment. To construct CP odd observables, we will use the lepton
momenta, which are directly measured in experiment and are related to $q_1$
and $q_2$ through a Lorentz boost. We denote these momenta with
$q_+=(E_+,{\bf q_+})
$ and $q_-=(E_-, {\bf q_-})$ for the lepton $\ell^+$ and $\ell^-$ repectivly.
We construct the following CP odd observables:
  $$ \eqalign { O_1 &= {E_+-E_- \over M_W} \cr
                O_2 &=({\bf \hat p}\cdot {\bf \hat q_+})^2
                     -({\bf \hat p}\cdot {\bf \hat q_-})^2 \cr
                O_3 &={\bf \hat p}\cdot ({\bf \hat q_+}-{\bf\hat q_-})
                      {\bf\hat p}\cdot ({\bf\hat q_+}\times {\bf\hat q_-})
      \cr} \eqno(2.10) $$
with
  $$ {\bf\hat q_+} ={{\bf q_+}\over \vert {\bf q_+}\vert},\
   {\bf\hat q_-} ={{\bf q_-}\over \vert {\bf q_-}\vert} \eqno(2.11) $$
and the vector ${\bf \hat p}$ is the direction of motion of the electron or
positron.
Because of the Bose symmetry of the two photon initial state the expectation
value of any observable which is odd in ${\bf\hat p}$, is zero. With these
observables one can also define the corresponding CP asymmetries:
  $$   A_i={ N(O_i >0) -N(O_i <0) \over N(O_i>0)+N(O_i<0)} \ \ (i=1,2,3)
   \eqno(2.12) $$
Where $N(O_i >0)$($N(O_i <0)$) denotes the number of events with $O_i >0$(
$O_i <0$).
Any nonzero $<O_i>$ or any nonzero$A_i$ indicates CP violation.
Further, the observables
$O_1$ and $O_2$ are CPT odd, the expectation values of them and the
corresponding asymmetries can be nonzero only if an absorbtive part of the
amplitude for (2.1) and CP violation exist. \par\vskip 30pt
\noindent {\bf 3. CP Violation in Two Higgs--Doublet Models}\par
In the process we are studying
the effect of CP violation from the minimal standard
model is zero up to two loop level at least and is thus too
 small to be observed.
We consider two Higgs--doublet extensions of the minimal standard model. In
these extensions CP violation is due to the complex expectation values
of the Higgs-doublets. However, in general CP violation
occours together with
the presence of flavour changing neutral currents at the tree level.
The flavour changing
neutral currents can be eliminated by imposing some discrete
symmetry on the Lagrangian. This discrete symmetry is
softly broken in the Higgs potential[5]. In this way it
is possible to construct models with
CP violation due to the complex expectation value of the Higgs doublets
in absence of the flavour changing neutral currents. \par
The effect of CP violation from these extensions has been
studied in top quark decay[6] and
in the $t\bar t$ system produced at $pp$ colliders[7] or at $e^+e^-$[8,9]
colliders, where CP asymmetries can be large as $10^{-3}$. CP violation
in the interactions between the gauge bosons is also studied in [10] and [11].
The process considered here gives
another oppertunity to study CP violation in such extensions.
\par
In the extensions mentioned above,
CP violation in the process discussed in Sect. 2 can be obtained at one loop
level. At this level there is only one type of diagram which introduces
CP violation. This type of diagram is shown in Fig. 1, where the loop
is a fermionic loop. CP violation is caused by the couplings between
$\bar f i\gamma_5 f$ and neutral Higgs fields, and the heaviest fermion
is dominant. In the following we will take only the top quark into account
and employ the notation used in [10] and [11]. In this notation CP violation
due to the neutral Higgs exchange
is paramerized with a $3\times 3$ real othogonal matrix $d$, the nonzero
off diagonal
matrix elements $d_{3j}$ and $d_{j3}$($j=1,2$) indicating CP violation. The
couplings involved in Fig.1 are:
$$ \eqalign { L =& e{m_t\over 2M_W\sin\theta _W}{\rm ctg}\beta\, d_{3j}
    \phi_j\bar t i
\gamma_5 t  -i{2\over 3}e\,\bar t \gamma_\mu t A^\mu \cr
    & +e{M_W \over \sin\theta_W}
     d_{1j}\phi_j W^+_{\mu}W^{-,\mu} \cr} \eqno(3.1)$$
where $\phi_j(j=1,2,3)$ are the mass eigenstates of the neutral Higgs fields,
${\rm  ctg}\beta=v_2/v_1$ is the ratio of the absolute expectation
values of the
two Higgs-doublets.
We assume that the $\phi_1$ is the lightest Higgs particle and is
dominant in Fig. 1. From (3.1) and Fig. 1 we obtain the CP violating amplitude
$T_A$:
  $$ \eqalign { T_A &= {16\over 3}{\alpha^2 \over \sin\theta_W^2}
     {\rm ctg}\beta\, d_{11}d_{31} I({\sqrt{\hat s} \over 2m_t}) D_H(\hat s,
    M_H) \cr
    & \cdot \varepsilon^{\nu_1}(p_1)\varepsilon^{\nu_2}(p_2)
      \varepsilon^*_{\mu_1}(k_1)\varepsilon^*_{\mu_2}(k_2) g^{\mu_1\mu_2}
      \varepsilon_{\nu_1\nu_2\alpha\beta}p_1^\alpha p^\beta_2 \cr}
   \eqno(3.2) $$
with
 $$ \eqalign { \hat s& =(p_1+p_2)^2, \cr
        I(z)&= \cases {
           {1\over 2z^2} ({\rm arcsin}( z))^2, \ \ {\rm for}\  z\le 1 \cr
           {1\over  2z^2} ({\pi \over 2} +{i\over 2} {\rm ln}
    { z+\sqrt{z^2-1} \over z-\sqrt{z^2-1}})^2, \ \ {\rm for}\  z>1 \cr}
   \cr }
   \eqno(3.3) $$
Here $M_H$ stands for the mass of $\phi_1$ and $D_H$ is its propagator
which is discussed further in Sect. 4. The CP violating part of the quantity
$R$ defined in (2.7) is then obtained through the interference
between $T_A$ and the amplitude for (2.1) at the tree-level from
the standard model. \par
In (3.2) the coupling parameters ${\rm ctg}\beta$ and $d_{11}d_{13}$ are
unknown. From the upper bound of the electric dipole moment of the neutron
one can not obtain enough information to constrain $d_{11}d_{13}$. From the
fact that the $d$ is a $3\times 3$ real othogonal matrix an upper bound
can be derived:
  $$ d_{11}d_{13} \le {1\over 2} \eqno(3.4) $$
As to the ratio ${\rm ctg} \beta$, certain discrete symmetries may lead
to the so called "natural choice", ${\rm ctg}\beta  < 1$, based on the
observation that $m_t >> m_b$. However, not all discrete symmetries, which
can be imposed on the theory to eliminate the flavour changing neutral
currents at tree level, lead to ${\rm ctg}\beta < 1$, for example, the
models I and III listed in [8]
allow ${\rm ctg}\beta >1$. It is possible that the ratio ${\rm ctg}\beta$
may be larger than one.
\par
It should be pointed out that CP violation in these
 extensions of the SM can also be
studied in the single Higgs production via photon fusion, if the
polarization of the initals photons is observed and the Higgs is
light enough to be produced. A detailed analysis in this case
was  made in [12]. \par\vskip 30pt
\noindent
{\bf 4. The numerical Results and the Summary} \par\vskip15pt
Before presenting our numerical results for the observables we defined
above we give
a detailed discussion on the Higgs progagator. As is well known,
the absorptive part in the propagator may lead to some signficant effect,
although the absorptive part usually comes from higher order. For
$M_H < 2M_W$ we neglect this part in our calculation since it can be
expected that this part is too small to give significant contributions
to our observables. Therefore, we take the propagator $D_H$ to have the form:
  $$ D_H(\hat s, M_H) = {1 \over \hat s -M_H^2}, \ \
 {\rm for}\ M_H< 2M_W \eqno(4.1) $$
In this case, the absorbtive(dispersive) part of the amplitude $T_A$ in (3.2)
corresponds to ${\rm Im}I(z) ( {\rm Re} I(z))$. \par
For $M_H >2 M_W$, the absorptive part in the $D_H$
does lead to significant effects
in our observables. We parametrize the propagator
in Breit--Wigner approximation
in term of the total decay width $\Gamma_H$:
   $$ D_H(\hat s, M_H)= { (\hat s-M_H^2) -i\Gamma_H M_H \over
           (\hat s -M_H^2)^2 +\Gamma_H^2 M_H^2 }, \ \
    {\rm for}\ M_H > 2M_W   \eqno(4.2) $$
and the absorptive(dispersive) part of the amplitude $T_A$ is corresponding
to ${\rm Im} (I(z)\cdot D_H)$
$({\rm Re} ( I(z)\cdot D_H))$.
Since the $\Gamma_H$ in the two Higgs-doublet extensions depends on the
unknown mass $M_H$ as well as the unknown coupling parameters $d_{ij}$
and ${\rm tg}\beta ({\rm ctg}\beta)$, our observables in general have
a complicated dependence on the unknown coupling parameters. However,
for the small $\Gamma_H$ the expression in (4.2)
can be approximated by:
    $$ D_H(\hat s, M_H) = {1 \over \hat s- M_H^2} -i\pi\delta (\hat s
 -M_H^2) \eqno(4.3) $$ \par
We used the both expressions for the $D_H$ in our numerical calculations
of our CP odd observables. By varying $\Gamma_H$ from $0$ to 40GeV we find
that the expression in (4.3) is a good approximation for our observables.
However, for $M_H > 2m_t$ $\Gamma_H$ can be very large because of
the large
mass $M_H$ and the new decay channel. For such large values of $\Gamma_H$,
for example, for $\Gamma_H > 100$GeV,
some numerical results of our observables can be changed in order of $50\%$
compared with them by small $\Gamma_H$. Keeping this in mind, we present
only our results calculated with the small $\Gamma_H$ and in this case the
expectation value of the CP odd observables or the CP asymmetries are
proportional to the product of $d_{11}d_{31}$ and ${\rm ctg}\beta$. \par
We take the photon distribution function given in [1], where we assume
that the laser energy is 1.26eV and the $e-\gamma$ conversion factor is one.
To simulate experimental cuts, we select for measuring the CP violation effect
only these events, in which the lepton energy $E_{+(-)}$ is larger
than 10GeV and the angle of the outgoing leptons with respect to the electron
or positron beam direction is not smaller than $10^0$. Using $\sqrt{s}=500$GeV
and $m_t=150$GeV we have for different Higgs mass $M_H$ the following
numerical results: \par
    For $M_H=100$GeV:
  $$ \eqalign { A_1 &=1.36\times 10^{-4} d_{11}d_{31}{\rm ctg}\beta, \ \ \
               <O_1> =1.46\times 10^{-4} d_{11}d_{31}{\rm ctg}\beta \cr
               A_3 &=1.68\times 10^{-4}d_{11}d_{31}{\rm ctg}\beta, \ \ \ \
                <O_3> =6.5\times 10^{-5} d_{11}d_{31}{\rm ctg}\beta \cr }
          \eqno(4.4) $$
\par
    For $M_H=200$GeV:
  $$ \eqalign { A_1 &=-2.82\times 10^{-4} d_{11}d_{31}{\rm ctg}\beta, \ \ \
               <O_1> =-3.1\times 10^{-5} d_{11}d_{31}{\rm ctg}\beta \cr
               A_3 &=2.35\times 10^{-4}d_{11}d_{31}{\rm ctg}\beta, \ \ \ \ \
                <O_3> =1.05\times 10^{-4} d_{11}d_{31}{\rm ctg}\beta \cr }
          \eqno(4.5) $$
\par
 For $M_H=350$GeV:
  $$ \eqalign { A_1 &=-7.65\times 10^{-4} d_{11}d_{31}{\rm ctg}\beta, \ \ \
               <O_1> =-7.36\times 10^{-4} d_{11}d_{31}{\rm ctg}\beta \cr
               A_3 &=3.45\times 10^{-4}d_{11}d_{31}{\rm ctg}\beta, \ \ \ \ \
                <O_3> =1.17\times 10^{-4} d_{11}d_{31}{\rm ctg}\beta \cr }
           \eqno(4.6) $$
\par
 For $M_H=500$GeV:
  $$ \eqalign { A_1 &=-1.35\times 10^{-4} d_{11}d_{31}{\rm ctg}\beta, \ \ \
               <O_1> =-1.53\times 10^{-4} d_{11}d_{31}{\rm ctg}\beta \cr
               A_3 &=-1.27\times 10^{-4}d_{11}d_{31}{\rm ctg}\beta, \ \ \
                <O_3> =-4.48\times 10^{-5} d_{11}d_{31}{\rm ctg}\beta \cr }
           \eqno(4.7) $$
\par
We do not present the results for $<O_2>$ and $A_2$ because they are one order
of magnitude
smaller than the results for $<O_1>$ and $A_1$. To determine the sensitivity of
our CP odd observables we also calculated the variances of them and the
cross section for the process (2.3) at NLC. Using the standard model at tree
level and taking the leptonic branching ratio($\approx 33\%$)
 of the $W$ decay into
account, we have under the conditions mentioned above:
   $$ \sigma =4{\rm pb},\ \ <O_1^2>=0.57, \ \
     <O_3^2>=0.096 \eqno(4.8) $$
Note that the variance for an asymmetry is identically 1.
Assumming the luminosity
per year at NLC to be $10$fb, the number of the available events is about
 $4\cdot 10^4$. We obtain then the statistical errors for our observables:
 $$ \delta A_1=\delta A_3 =\sqrt {1 \over N_{event}}=0.5\%, \ \ \
     \delta O_1=\sqrt { <O_1^2> \over N_{event}}=0.4\%, \ \ \
   \delta O_3=\sqrt { <O_3^2> \over N_{event}}=0.15\% \eqno(4.9) $$
\par
CP violation is detectable only if the $<O_i>$ or $A_i(i=1,2,3)$ are at least
larger than their statistical error. Taking $<O_1>$ at $M_H=350$GeV as an
example, the product $d_{11}d_{31}{\rm ctg}\beta$ should be larger than 5.4.
\par
To summarize: in this work we studied the possibility of detecting
 CP violation
in $\gamma\gamma \rightarrow W^+W^-$ at NLC, our CP odd observables and the
corresponding CP asymmetries are constructed with the directly measured
energies and momenta of the leptons from the $W$ decay. For the observables
we propose one can detect CP
violation without requiring  complete knowledge about the c.m.s.
of the initial photons and about the rest frame of the $W$ bosons.
Therefore,  our observables are easy to measure. The prediction
of the CP violation effects is worked out for two Higgs doublet
models. The effect of the Higgs width is studied  and it is
significant.
If the Higgs sector of these models is not CP invariant, then CP
violation can be measured with our CP odd observables in some parameter
region. \par\vskip 20pt
 {\bf Acknowledgment:}\par
We thank Dr. X. G. He and Dr. S. Tovey for useful discussions. \par
\vfil\eject
\centerline{\bf References}\par\vskip 30pt\noindent
[1] I.F. Ginzburg et al., Nucl. Instrum. Methods. 205 (1983) 47 \par\noindent
[2] S.Y. Choi and F. Schrempp, Phys. Lett. B272 (1991) 149 \par\noindent
[3] E. Yehudai, Phys. Rev. D44 (1991) 3434 \par\noindent
[4] M. Kobayashi and T. Maskawa, Prog. of Theo. Phys. 49 (1973) 652
 \par\noindent
[5] G.C. Branco and M.N. Rebelo, Phys. Lett. B160 (1985) 117 \par\noindent
\ \ \ \ J. Liu and L. Wolfenstein, Nucl. Phys. B289 (1987) 1 \par\noindent
\ \ \ \ S. Weinberg, Phys. Rev. D42 (1990) 860 \par\noindent
[6] B. Grzadkowski and J.F. Gunion, Phys. Lett. B287 (1992) 237 \par\noindent
[7] C.R. Schmidt and M. Peskin, Phys. Rev. Lett. 69 (1992) 410 \par\noindent
[8] W. Bernreuther, T. Schr\"oder and T.N. Pham, Phys. Lett. B279 (1992)
389 \par\noindent
[9] W. Bernreuther, O. Nachtmann, P. Overmann and T. Schr\"oder,
 Nucl. Phys. B388
\par\noindent \ \ \ \ (1992) 53\par\noindent
[10] A. M\'endez and P. Pomarol, Phys. Lett. B272 (1991) 313 \par\noindent
[11] X.G. He and J.P. Ma and B.H.J. McKellar, Melbourne--Preprint,
UM--P--92/25, \par\noindent
\ \ \ \ \  to be published in Phys. Lett. B \par\noindent
[12] B. Grzadkowski and J.F. Gunion, Phys. Lett. B294 (1992) 361
\par\noindent
\vfil\eject
\centerline{ Figure Caption }\par\vskip20pt
Fig.1. One of the two Feynman graphs for the CP violating amplitude.
The other one is to obtain through interchanging the two photons. \par
\vfil\eject\end